\documentclass{article}
\usepackage{emulateapj,pstricks,apjfonts}

\righthead{KEMPNER \& SARAZIN}
\lefthead{THE RADIO AND X-RAY LUMINOSITY OF ABELL~2199}

\submitted{Accepted by the Astrophysical Journal.}

\begin{document}

\title{Limits on the Diffuse Radio and Hard X-ray Emission of Abell~2199}

\author{Joshua C. Kempner and Craig L. Sarazin}

\affil{Department of Astronomy, University of Virginia, 
P. O. Box 3818, Charlottesville, VA 22903-0818;
jck7k@virginia.edu,
cls7i@virginia.edu}

\begin{abstract}
The Westerbork Northern Sky Survey (WENSS) and the NRAO/VLA Sky Survey (NVSS)
were used to determine an upper limit to the diffuse radio flux from the
nearby cluster Abell~2199.
For the entire cluster, this limit is $<$3.25 Jy at 327 MHz from WENSS;
for the inner 15\arcmin\ radius, the limit is
$<$168 mJy at 1.4 GHz.
These limits are used to constrain the cluster magnetic field by
requiring that the radio flux be consistent with
the hard X-ray (HXR) flux observed by {\it BeppoSAX}, assuming that the
observed HXR excess is due to inverse Compton (IC) scattering of cosmic
microwave background photons by relativistic electrons in the intracluster
gas.
We find that the magnetic field must be very weak ($<$0.073 $\mu$G)
in order to avoid producing an observable radio halo.
We also consider the possibility that the HXR excess is due to nonthermal
bremsstrahlung (NTB) by a population of suprathermal electrons which are
being accelerated to higher energies.
We find that a NTB model based on a power-law electron momentum
distribution with an exponent of $\mu \approx 3.3$ and containing about
5\% of the number of electrons in the thermal ICM can reproduce the
observed HXR flux.
\end{abstract}


\section{Introduction} \label{sec:intro}

Abell~2199 is a nearby ($z=0.0303$) X-ray cluster.  It contains a
cooling flow which is centered on the cD galaxy NGC~6166, which is also
a strong radio source (3C~338).
Recent X-ray observations of the cluster with {\it BeppoSAX}
(Kaastra, Bleeker, \& Mewe 1998;
Kaastra et al.\ 1999) found
evidence for an excess of hard X-rays above that expected from the
thermal emission of the hot intracluster gas.
There are several possible interpretations of this hard tail to the X-ray
luminosity.
One is that the excess is caused by inverse Compton (IC) scattering of cosmic
microwave background (CMB) photons by highly relativistic cosmic ray electrons
located in the intracluster medium (ICM)
(Rephaeli 1979;
Kaastra et al.\ 1999).
Another is that the excess is produced by
nonthermal bremsstrahlung (NTB) by subrelativistic but suprathermal electrons
which are undergoing acceleration in the ICM
(Kaastra et al.\ 1998;
En{\ss}lin, Lieu, \& Biermann 1999;
Sarazin \& Kempner 1999).
If the first explanation
is correct, the same relativistic electrons should produce diffuse radio
synchrotron emission as long as the ICM contains a magnetic field which
is not too small.
As a test of the IC model, we search for diffuse
radio emission from this cluster using the Westerbork Northern Sky
Survey (WENSS; Rengelink et al.\ 1997) and the NRAO/VLA Sky Survey
(NVSS; Condon et al.\ 1998).
We assume $H_0 = 50$ km s$^{-1}$ Mpc$^{-1}$ and $q_0 = 0.5$ throughout.

\section{Limits on Diffuse Radio Halo} \label{sec:radio}

We searched for diffuse radio emission from Abell~2199 using the WENSS
92~cm radio survey and the NVSS radio survey at 20~cm.  The WENSS images
have a synthesized beam of 54\arcsec$\times$85\arcsec\ at the
declination of Abell~2199.  The typical noise level is 3.6~mJy
beam$^{-1}$.  WENSS interferometer observations should be sensitive to
radio emission on scales of less than $\sim$1$^{\circ}$.  The NVSS synthesized
beam is 45\arcsec$\times$45\arcsec.  The typical noise level is 0.45~mJy
beam$^{-1}$.  The NVSS is sensitive to emission on scales of less than
15\arcmin.  At the redshift of Abell~2199, a typical cluster core radius
of 200~kpc corresponds to 3\farcm5, while an Abell radius of
3~Mpc subtends 58\arcmin.  Thus, the WENSS observations should be
sensitive to a diffuse radio halo in the cluster if it is smaller than
the cluster Abell radius, while the NVSS should be sensitive to a halo
smaller than 775~kpc.

The WENSS survey gives a source catalog (Rengelink et al.\ 1997), which
contains 22 sources which lie within half a degree (the largest angular
scale not resolved out by the interferometer used in the survey) of the
center of the cluster.  We adopt the position of the central cD galaxy
NGC~6166, R.A.\ = 16$^{\rm h}$26$^{\rm m}$55\fs4,
Dec.\ = +39\arcdeg39\arcmin37\arcsec, as the center of the cluster.
The strongest radio source is 3C~388 which is associated with the central cD
galaxy.  None of these sources is extended on the scale of the cluster,
so we do not include this emission in the measurement of the diffuse
cluster radio emission.  To search for diffuse radio emission, we used
the survey radio image to determine the total radio flux from a circle
of radius 30\arcmin\ (1.55~Mpc at the distance of Abell~2199).  We then
subtracted from this measured flux the sum of the integrated fluxes of
all sources in the WENSS catalog within this region.  This residual flux
was found to be $3.25 \pm 0.21$~Jy.

However, the cD galaxy in Abell~2199 is the very strong radio source
3C~338, with a peak-to-rms brightness ratio of 3190 in WENSS.  Slight
instrumental variations between receivers in a radio interferometer
cause the region immediately around a bright source with such a large
peak-to-rms ratio to have both an increased rms noise level and a
systematic positive flux from uncleaned sidelobes in the absence of real
diffuse emission
(Condon et al.\ 1998).
This systematic excess flux varies from source to source in WENSS.  We
did an inspection of several other bright sources in the surveys and
found that the effect is consistently between about 8 and 12\% of the
peak intensity of the source.
The residual flux in Abell~2199 is approximately 12\% of the
peak flux of the cD, so we must treat this residual as an upper
limit on the diffuse synchrotron flux from the cluster rather than as a
detection of diffuse emission.
If we subtract our best estimate of this systematic effect from the
measured flux, the upper limit on the flux would be reduced by at least
a factor of five.  However, due to the uncertainty in the magnitude of
this baseline offset, we use the more conservative value without
attempting to remove the effect of the offset baseline.
Thus, our upper limit to the diffuse radio flux of Abell~2199 at 92 cm is
3.25 Jy.

The dynamic range of the NVSS is more sensitive, and the cD is better
resolved than in the WENSS observations, so the peak-to-rms of the cD is
not as great. There is still some noticeable uncleaned sidelobe
structure in the image, but it is relatively minor.  Since the NVSS
resolves out structures larger than 15\arcmin, we used the same
procedure as with WENSS, but restricted to a 15\arcmin circle.  We
measure a residual flux of $-97\pm56$~mJy.
The small negative flux appears to be caused by the uncleaned sidelobes
and is consistent
with zero diffuse emission. From this we can rule out the existence of a
radio halo within the central 750~kpc of the cluster.
The 3-$\sigma$ upper limit of the diffuse flux in the NVSS survey is
$<$168~mJy.
The WENSS limit is more conservative since, unlike the NVSS, no flux is
resolved out in the region containing the HXR emission observed by
{\it BeppoSAX}.

\section{IC X-ray Emission and the Cluster B Field} \label{sec:IC}

If the relativistic electrons in the cluster have a power-law energy
distribution, then both the radio synchrotron and the IC hard X-ray
spectra are expected to be power laws in frequency with the same
spectral index
(e.g., Rephaeli 1979).
Kaastra et al.\ (1998, 1999) fit the X-ray spectra from {\it BeppoSAX}
with a power-law spectrum excess to the thermal emission from the
ICM.
In Kaastra et al.\ (1999), a single power-law excess was fit across
the entire X-ray band, roughly from 0.1--100 keV.
The total luminosity in this band was
$(1.30 \pm 0.32) \times 10^{44}$ ergs s$^{-1}$.
The cluster Abell~2199 shows both a hard X-ray excess and and extreme
ultraviolet (EUV) or soft X-ray excess
(Lieu, Bonamente, \& Mittaz 1999).
In Kaastra et al.\ (1999), the authors argue that both the EUV and HXR
excesses are nonthermal IC emission.
The best-fit power-law photon spectral index was
$\Gamma = 1.81 \pm 0.25$ for the entire spectral band 0.1--100 keV.
This implies a flux of
$S_{HXR} = (1.6 \pm 0.4) \times 10^{-11}$ ergs cm$^{-2}$ s$^{-1}$
in the HXR band, 10--100 keV.

If this hard X-ray emission is due to IC scattering of the
CMB photons, then the predicted radio synchrotron flux is
\begin{eqnarray}
S_{\nu} & = & 234 \,
\left(\frac{S_{HXR}}{10^{-11} \, {\rm ergs \, cm^{-2} \, s^{-1}}} \right) \,
\left(\frac{B}{1 \, \mu{\rm G}} \right)^{1.81}
\nonumber \\
& & \qquad \times \left(\frac{\nu}{327  \, {\rm MHz}} \right)^{-0.81}
\, {\rm Jy} \, ,
\label{eq:radio2}
\end{eqnarray}
if the photon spectral index of the HXR emission is $\Gamma = 1.81$.
Here, $S_{HXR}$ is the HXR flux in the 10--100 keV band,
$B$ is the cluster magnetic field,
and $\nu$ is the observed frequency.

If we use the observed HXR flux of Abell 2199 and our conservative
upper limit on the radio flux at 327 MHz ($S_\nu < 3.25$ Jy), 
this implies a strong upper limit on the magnetic field of
$B < 0.073$ $\mu$G.
If we correct for the aforementioned instrumental excess radio emission
near strong sources, we reduce the observed radio flux by a factor of
$\sim8$ and consequently reduce the maximum magnetic field by a factor
of 2 to 3.
Similarly, if the NVSS limit on a smaller radio halo at higher frequencies
is used, the limit on the magnetic field is
$B \la 0.02$ $\mu$G.
Thus, we have a conservative upper limit of 0.073 $\mu$G, with the strong
suggestion that the field must be even weaker.
If we instead assume a more typical magnetic field of 1 $\mu$G, the
upper limit on the radio flux implies an upper limit to the inverse Compton
HXR flux which is more than 100 times fainter than was observed with
{\it BeppoSAX}.

\section{Nonthermal Bremsstrahlung} \label{sec:bremss}

Alternatively, the HXR excess observed in Abell~2199 might result from
some other emission process.
Perhaps the most likely possibility is nonthermal bremsstrahlung (NTB)
from suprathermal electrons in the ICM
(Kaastra et al.\ 1998;
En{\ss}lin, Lieu, \& Biermann 1999;
Sarazin \& Kempner 1999).
These might be electrons with energies $\ga 10$ keV which are currently
being accelerated up to much higher energies, either by shocks or by
turbulent acceleration.
Detailed models for NTB emission in clusters are given in Sarazin \&
Kempner (1999).

One feature of such models is that the excess emission spectrum should
flatten at low energies, because the suprathermal population only contains
electrons with energies which are higher than typical thermal energies
(see Figure~\ref{fig:ntb} below).
Thus, one does not expect nonthermal bremsstrahlung to produce any
EUV excess emission directly.
It is also inappropriate to fit a single power-law spectrum to the NTB
emission across the entire X-ray band 0.1--100 keV.
On the other hand, Sarazin \& Kempner (1999) show that a power-law does
provide a reasonable fit to the spectrum of the HXR emission,
10--100 keV.
In Kaastra et al.\ (1998), the HXR excess emission in Abell~2199
as seen in the {\it BeppoSAX} PDS was fit independently of any softer excess.
This gave a flux of $S_{HXR} = (1.4 \pm 0.4) \times 10^{-11}$ ergs
cm$^{-2}$ s$^{-1}$ in the 10--100 keV band.
The best-fit power-law photon spectral index was
$\Gamma = 2.5^{+1.1}_{-0.8}$.
This is steeper than the spectral index found by fitting a power-law excess
to the entire X-ray band
(Kaastra et al.\ 1999),
although the error bars overlap.

Let $N ( p ) d p$ be the total number of nonthermal electrons with
normalized momenta in the range $p$ to $p + dp$, where $p$ is the
momentum normalized to $m_e c$.
The simplest models for the nonthermal electrons in clusters have
a power-law momentum distributions
(Sarazin \& Kempner 1999),
with
\begin{equation} \label{eq:powerlaw}
N ( p ) = N_o p^{- \mu} \, \qquad p \ge p_l
\, .
\end{equation}
We will assume that the suprathermal population consists only of
particles with momenta $ p > p_l$, such that their kinetic energies
exceed $3 k T$ where $T$ is the temperature of the thermal ICM.
If the cooling flow at the center of Abell~2199 is excluded, the mean
ICM temperature is $k T = 4.8 \pm 0.2$ keV
(Markevitch et al.\ 1999b).
This implies that $p_l = 0.24$.

For steep power-law momentum distributions ($\mu \ga 3.5$), the
nonthermal bremsstrahlung HXR emission is nearly a power-law with
$\Gamma \approx 1 + \mu / 2$
(Sarazin \& Kempner 1999).
For flatter momentum distributions, the NTB spectra are still approximately
fit by power-laws, but the exponent is flatter than given by this
expression.
The observed spectral index of $\Gamma = 2.5$ between 10 and 100 keV is
produced by a model with $\mu = 3.33$.
The predicted nonthermal bremsstrahlung spectrum of this models is
shown in Figure~\ref{fig:ntb}.
The observed flux in the 10-100 keV band is reproduced by a model
with $N_o = 3.88 \times 10^{68}$, which implies that the total
number of nonthermal electrons is $4.6 \times 10^{69}$ if the electron
spectrum extends to high energies.
This represents about 5\% of the total number of thermal electrons
in the intracluster medium in Abell~2199
(Mohr, Mathiesen, \& Evrard 1999).

If the nonthermal electron distribution extends to much higher energies,
HXR would be produced through IC by these higher energy electrons.
In fact, if the electron spectrum is flatter than $\mu \la 2.7$, more
HXR emission is produced by IC than by NTB.
However, for the steep electron spectrum in our NTB model, IC by
high energy electrons only contributes about 0.6\% of the flux in
the 10--100 keV HXR band.
The small number of high energy electrons in this steep spectrum
model also reduces the radio synchrotron emission.
For a magnetic field of 1 $\mu$G, the predicted radio flux is
0.73 Jy at 327 MHz.
Thus, the predicted radio emission is consistent with our limit
as long as $B \la 2$ $\mu$G.

\centerline{\null}
\vskip2.55truein
\includegraphics{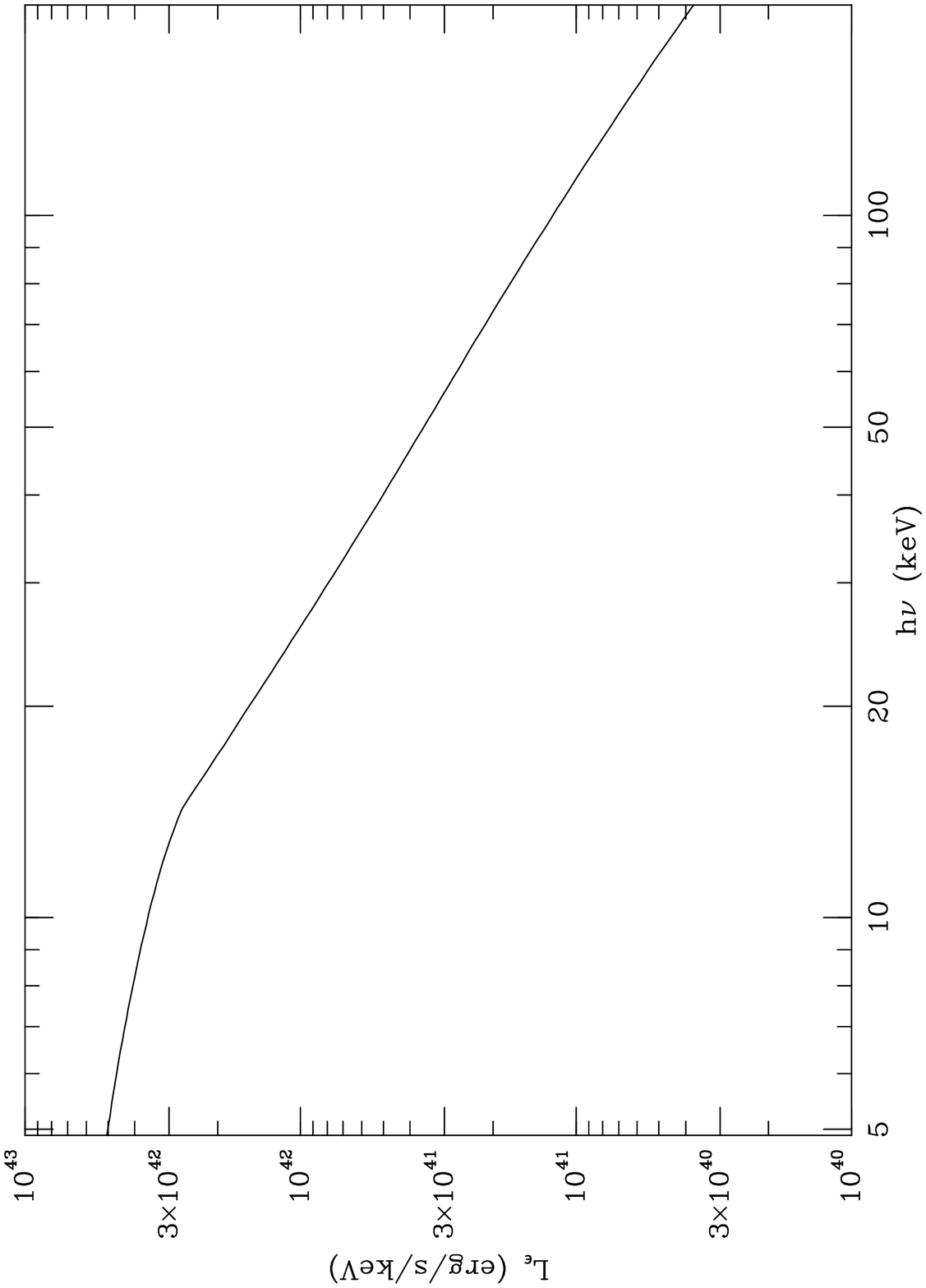}
\figcaption{The nonthermal bremsstrahlung hard X-ray emission of our
model with a power-law momentum distribution with $\mu = 3.33$.
The emitted spectrum is given as a function photon energy.
The flattening of the spectrum at low energies is due to the cut-off
in the suprathermal electron population at an energy of $3 k T$.
\label{fig:ntb}}

\vskip0.2truein

On the other hand, with this steep electron spectrum, the model
cannot reproduce the EUV or soft X-ray excess which has been
observed in Abell~2199
(Kaastra et al.\ 1999;
Lieu, Bonamente, \& Mittaz 1999),
either by NTB or IC emission.
Apparently, a distinct population is needed to produce the observed EUV
emission.
Since the electrons which generate EUV by IC have long lifetimes
comparable to cluster ages
(Sarazin \& Lieu 1998),
it is possible that the EUV emission is produced by
an older population of electrons, while the NTB HXR emission
is due to electrons currently being accelerated.

\section{Conclusions} \label{sec:conclude}

We have used the WENSS and NVSS radio surveys to search for any diffuse
radio emission (a radio halo or relic) associated with the cluster of
galaxies Abell 2199.
We do not detect any such emission.
The best limit on any cluster-wide emission comes from the WENSS survey,
which gives a limit of 3.25 Jy at 327 MHz.
The limit would be considerably tighter were it not for the confusing
effects of uncleaned sidelobes from the central radio source 3C~338.
The NVSS survey gives a much tighter limit of 168 mJy at 1.4 GHz, but
only applies to a centrally condensed radio halo with a size of less
than about 750 kpc.

The absence of diffuse radio emission in Abell~2199 is not surprising
when this cluster is compared to other clusters with and without radio
halos.
Abell~2199 is an extremely regular cluster with a strong central cooling
flow
(e.g., Markevitch et al.\ 1999b).
Radio halos and relics are relatively rare objects which are
generally associated with irregular clusters which are undergoing mergers
(e.g., Giovannini, Tordi, \& Feretti 1999).
It has generally been argued that the presence of diffuse radio emission
in these irregular clusters is the result of the acceleration or transport
of relativistic particles by shocks or turbulence associated with the
merger.
The presence of a strong cooling flow in Abell~2199 is also an indication
that this cluster has not had a recent strong merger, as cooling flows
and irregular cluster structures tend to be anticorrelated
(Buote \& Tsai 1996).
Clusters with radio halos also tend to be rather hot clusters
(Giovannini et al.\ 1999),
whereas Abell~2199 is fairly cool ($k T = 4.8 \pm 0.2$ keV;
Markevitch et al.\ 1999b), even when the temperature is corrected for the
cooling flow.

On the other hand, extended nonthermal hard X-ray emission was detected from
Abell~2199 with {\it BeppoSAX}
(Kaastra et al.\ 1998, 1999).
HXR tails in the spectra of clusters with radio halos are expected as a
result of IC scattering by high energy ($\ga$ 1 GeV) relativistic electrons
(Rephaeli 1979), and this is an explanation which has been proposed for
the HXR emission in Abell~2199
(Kaastra et al.\ 1999).
Unless the ICM magnetic field is very weak, the same electrons would 
produce diffuse radio synchrotron emission.
The absence of such diffuse radio emission in Abell~2199 is a significant
problem for the IC model for the HXR emission.

If we accept inverse Compton radiation as the explanation of the
observed hard X-ray tail, then the cluster magnetic field implied by
the limit on the radio flux is very weak.
If we adapt our most conservative limit on the total diffuse radio flux
of the cluster, it implies that the ICM magnetic field is
$B < 0.073$ $\mu$G.
If we correct the radio limit for the uncleaned sidelobes of the
central radio source 3C~338, or if we use the stronger limit on
a centrally condensed radio halo ($\la 750$ kpc in diameter) from the NVSS,
then the limit on the magnetic field is smaller by a factor of about 3.
These field limits are at least an order of magnitude smaller than the
magnetic fields derived from Faraday rotation measurement towards individual
radio galaxies
(e.g., Feretti et al.\ 1995)
or statistical samples of radio sources
(Clarke 1999)
in other clusters.
Very strong Faraday rotation is detected toward the central
radio source 3C~338 in Abell~2199
(Ge \& Owen 1994),
which implies the presence of a magnetic field which
is about two orders of magnitude stronger than the limit we find for the
diffuse field if the HXR emission is due to IC scattering.
However, the field around 3C~338 might have been enhanced by compression
or shear associated with the cooling flow at the center of the cluster
(Soker \& Sarazin 1990).

One possibility is that the magnetic field in Abell~2199 (and, presumably,
other clusters) is very inhomogeneous, and the magnetic field and
relativistic electrons are anticorrelated
(En{\ss}lin et al.\ 1999).
This might occur because electrons in high magnetic field regions lose
energy rapidly by synchrotron emission,
and the remaining high energy electrons might be found preferentially in
weak magnetic field regions.
This model was proposed for the Coma cluster
(En{\ss}lin et al.\ 1999),
where the observed IC HXR emission implies a lower value of the magnetic
field than has been determined from Faraday rotation measurements
(Fusco-Femiano et al.\ 1999).

If the intracluster magnetic field and relativistic electrons are not
anticorrelated, it is difficult to believe that the ICM magnetic field
is as weak as required by the HXR flux.
Of course, it is possible that the nonthermal HXR flux is in error, 
because of calibration uncertainties with {\it BeppoSAX},
or because of another hard X-ray source in the field of view, or
because of a complex thermal structure in the ICM which produces a
thermal HXR tail.

Alternatively, the HXR emission may be real, but not due to IC scattering
of CMB photons.
The most attractive alternative emission mechanism may the nonthermal
bremsstrahlung emission by a population of mildly subrelativistic nonthermal
electrons with energies of 10-1000 keV
(Kaastra et al.\ 1998;
En{\ss}lin, Lieu, \& Biermann 1999;
Sarazin \& Kempner 1999).
We constructed a NTB model for the HXR emission in Abell~2199, based on
a power-law momentum distribution for the nonthermal electrons starting
at an energy of $3 k T$.
A model with a power-law exponent of $\mu = 3.33$ containing about 5\% of
the thermal electron population of the cluster fits the {\it BeppoSAX}
observations acceptably.
With a power-law exponent this steep, the electron population can be
extended to arbitrarily high electron energies without producing too much
HXR IC emission or radio synchrotron emission.
Assuming the same electron spectrum extrapolates to very high energies,
our limit on the diffuse radio emission is consistent with any magnetic
field with an average value of $\la 2$ $\mu$G.

This steep nonthermal electron population could not produce the observed
EUV and soft X-ray excess observed from Abell~2199
(Kaastra et al.\ 1999;
Lieu et al.\ 1999)
either by NTB or IC emission.
A distinct electron population is needed to produce this emission.
The electrons which would emit EUV by IC have lifetimes which are
nearly comparable to cluster ages
(Sarazin \& Lieu 1998).
Thus, it is possible that the EUV is generated by
an older population of electrons, while the NTB HXR emission
is due to electrons currently being accelerated out of the thermal
population.

The most likely source of the nonthermal subrelativistic electrons would
be acceleration of electrons out of the thermal distribution.
The acceleration might be due to shocks or turbulence
(e.g., plasma waves).
The required electron spectrum is about one power steeper ($\mu \approx 3.3$)
than is normally associated with particle acceleration in strong shocks
($\mu \approx 2$).
However, in simple shock models, the exponent of the power-law depends on
the shock compression $r$ as
$\mu = ( r + 2 ) / ( r - 1 )$
(e.g., Bell 1978).
In shocks in the intracluster gas, the preshock gas is quite hot, and the 
Mach numbers are expected to be several, rather than being very large.
The NTB model electron exponent of $\mu \approx 3.3$ corresponds to a shock
compression of $r \approx 2.3$ and a Mach number ${\cal M} \approx 2$.
These values are similar to those derived from the thermal and density
structures in several merging clusters
(Markevitch, Sarazin, \& Vikhlinin 1999a).
Alternatively, the electron acceleration might be due to turbulence.
This could give a steep power-law spectrum if the acceleration were
relatively inefficient.

One concern with the NTB model is that Abell~2199 is very regular, and
shows no real evidence for any hydrodynamical activity which might
generate shocks or turbulence.
On the other hand, the required particle spectrum is fairly steep, and
might be produced by relatively weak shock or turbulent acceleration, which
might persist for long periods after any major dynamical event.
Such weak shocks and turbulence might even be continually generated by the
infall of small galaxy groups into the clusters or possibly by the motions
of galaxies.
If so, then one would expect to find comparable nonthermal HXR tails in
most clusters of galaxies.

\acknowledgments
This work was supported in part by NASA Astrophysical Theory Program grant
NAG 5-3057.

\end{document}